\documentclass{JHEP3}
\usepackage{graphicx}
\usepackage{amsmath}

\newcommand{\MeV}{\text{\,MeV}}

\newcommand{\be}{\begin{equation}}
\newcommand{\ee}{\end{equation}}
\newcommand{\ba}{\begin{eqnarray}}
\newcommand{\ea}{\end{eqnarray}}
\renewcommand{\d}{\partial}
\renewcommand{\l}{\left(}
\renewcommand{\r}{\right)}

\newcommand{\e}{\mathrm{e}}

\renewcommand{\L}{\mathcal{L}}
\newcommand{\half}{\frac{1}{2}}
\newcommand{\dm}{\partial_\mu}

\preprint{RBRC 1012}

\title{Light inflaton after LHC8 and WMAP9 results}

\author{F. Bezrukov\\
  Physics Department, University of Connecticut, Storrs, CT
  06269-3046, USA and\\
  RIKEN-BNL Research Center, Brookhaven National Laboratory, Upton,
  NY 11973-5000, USA\\
  E-mail: \email{Fedor.Bezrukov@uconn.edu}}
\author{D. Gorbunov\\
  Institute for Nuclear Research of the Russian Academy of Sciences, 
  Moscow 117312, Russia and\\
  Moscow Institute of Physics and Technology, 
  Dolgoprudny 141700, Russia\\ 
  E-mail: \email{gorby@ms2.inr.ac.ru}
}

\abstract{We update the allowed parameter space of the simple chaotic
  inflationary model with quartic potential and light inflaton
  \cite{Bezrukov:2009yw} taking into account recent results from
  cosmology (CMB observations from SPT, ACT and WMAP) and from
  particle physics (LHC hints of the SM Higgs boson).  The non-minimal
  (yet small) coupling to gravity of the inflaton becomes essential to
  fit the observational data.  The inflaton has mass above 300\,MeV
  and can be searched for at $B$-factories in $B$-meson two-body
  decays to kaon and inflaton.  The inflaton lifetime depends on the
  model parameters, resulting in various inflaton signatures: either a
  missing energy, or a displaced vertex from the $B$-meson decay
  position, or a resonance in the Dalitz plot of a three particle
  decay.  We also discuss the implementation of the inflaton model to
  the $\nu$MSM, where the inflaton can be responsible for production
  of the dark matter sterile neutrino in the early Universe.}

  
\begin{document}

\section{Introduction}
\label{sec:intro}

Elegant solutions to the major problems of the Hot Big Bang theory are
provided by a preliminary inflationary stage of the Universe expansion
\cite{Starobinsky:1980te,Mukhanov:1981xt,Guth:1980zm,
  Linde:1981mu,Albrecht:1982wi}.  The simplest realization is made
using a scalar degree of freedom, inflaton, which evolves very slowly
at the inflationary stage providing nearly constant energy density and 
hence the accelerated expansion.  At later times the inflaton starts to
oscillate and the Universe exits inflation and enters a stage when
rapidly evolving homogeneous inflaton dominates the total energy
density.  The energy has to be transferred to the Standard Model (SM)
particles thus reheating the Universe.  To make the transfer an
inflaton-to-SM coupling is needed.  This coupling takes care of the SM
particle production and the onset of Hot stage of the Universe
expansion.

Let us analyse the consequences of adding explicitly just one scalar
inflaton field to the SM.  In a renormalizable model of the inflaton
sector the coupling to the SM should be renormalizable as well.  Then
for the simplest inflationary models with only one scalar field,
inflaton, involved in the inflationary dynamics, the only suitable
coupling to the SM is its coupling to the SM Higgs field via the
scalar potential.  Then reheating happens via the Higgs boson
production by oscillating inflaton field.  A variant of these models
has been considered in
Refs.~\cite{Shaposhnikov:2006xi,Bezrukov:2009yw}.  It is distinguished
by the absence of dimensionful parameters in the sector of SM fields.
The vacuum expectation value of the SM Higgs field $v$ appears as a
nonzero vacuum expectation value of the inflaton field: the quartic
coupling between the two field is responsible for both reheating in
the post-inflationary Universe and electroweak phase transition later
at the Hot stage.  The only dimensionful parameter is in the inflaton
sector, and it is of the order of electroweak scale.  All model
parameters are therefore stable with respect to the finite quantum
corrections (or logarithmic renormalization running).

The appealing phenomenological feature of the inflationary model
described above is the new light particle, inflaton, with the light
mass in GeV region.  Cosmology limits the mass range of the light
inflaton from above and from below: lighter inflaton causes very late
reheating, while quantum corrections from the inflaton-to-Higgs
quartic coupling spoil flatness of the inflaton potential for heavy
inflaton masses.  The latter bound on the inflaton-to-SM coupling
constrains the inflaton mass to be below 2\,GeV
\cite{Bezrukov:2009yw}.

In the model the quartic coupling induces the Higgs-inflaton mixing,
so the inflaton mass is proportional to the mass of the Higgs boson
and square root of the quartic coupling.  The inflaton mass range
presented above refers to the Higgs mass of 100--200\,GeV.  Recent
results from LHC \cite{:2012gk,:2012gu} can be interpreted as evidence
for the SM 125\,GeV Higgs boson, which allows to resolve the ambiguity
due to the Higgs mass value: now the only free parameter remaining in
the model is the Higgs-inflaton quartic coupling, or equivalently the
inflaton mass.  Thus the estimates given in
Ref.~\cite{Bezrukov:2009yw} can be appropriately refined.  It is
important for the phenomenology of the light inflaton, studied there.
Indeed, provided mixing with the SM Higgs boson, inflaton may be
produced in scatterings and decays of the SM particles and
subsequently decays into the SM particles.  The latter branching
rations are the same as those for a (hypothetical) SM Higgs boson of
mass equal to those of the inflaton.  Thus, \emph{the inflaton mass
  alone determines the inflaton production rate in the laboratory
  experiment and its lifetime,} which was estimated as
$10^{-8}$--$10^{-10}$\,s.  The most promising place to search for the
inflaton was suggested to be in $B$-meson decays
\cite{Bezrukov:2009yw}.

Another issue has risen due to the recent analyses of cosmological
data, mainly observations of the Cosmic Microwave Background (CMB)
anisotropy by SPT~\cite{Story:2012wx}, ACT~\cite{Sievers:2013wk}, and
the common global fit together with the analysis of CMB data collected
for 9 years by WMAP~\cite{Hinshaw:2012fq}.  The most important,
relevant for our model result is lowering an upper limit on the
tensor-to-scalar perturbations ratio, which excludes the simple
chaotic inflation with quartic potential discussed above at more than
95\% confidence level.  It was already pointed in
Ref.~\cite{Bezrukov:2009yw} that with a small non-minimal coupling to
gravity the situation may be cured.  It was
found~\cite{Bezrukov:2009yw} that, at rather small value of
non-minimal coupling constant $\lesssim10^{-3}$, this also leads to
the increase of the inflaton self-coupling constant.  The latter is
determined from the amplitude of the scalar perturbation power
spectrum obtained mostly from analyses of CMB anisotropy.  This change
was accounted for in the estimates presented in
Ref.~\cite{Bezrukov:2009yw}.  However, with new results from CMB
anisotropy \cite{Story:2012wx,Hinshaw:2012fq,Sievers:2013wk} a
somewhat larger non-minimal coupling is needed to make the inflaton
model viable, compared to the value adopted in
Ref.~\cite{Bezrukov:2009yw}.  Hence that analysis has to be modified
appropriately.

In the present paper we update the analysis of
Ref.~\cite{Bezrukov:2009yw} to account for both recent measurement of
the Higgs boson mass of about 126\,GeV \cite{:2012gk,:2012gu} and
recent new upper limit on the tensor-to-scalar ratio $r<0.13$ (at
95\%~C.L.~\cite{Hinshaw:2012fq}).

The paper is organized as follows.  In Sec.~\ref{sec:Model} we present
the model Lagrangian and discuss the particle spectrum in the scalar
sector.  In Section \ref{sec:Cosmo}, which is devoted to inflationary
dynamics of the model, we give predictions for cosmological parameters
to be tested in future cosmological experiments.  Particle physics
phenomenology is discussed in Sec.~\ref{sec:Pheno}, where we refine
the estimates of the mass range of the light inflaton and discuss the
strategy of searches for light inflaton in particle physics
experiments, such as LHCb and Belle-II.  Section~\ref{sec:Conclusion}
contains conclusions and further discussion on the model extension to
solve the major phenomenological problems of the SM: neutrino
oscillations, dark matter phenomenon and baryon asymmetry of the
Universe; here an example is
$\nu$MSM~\cite{Asaka:2005an,Asaka:2005pn}.

\section{Model: inflaton with $\beta X^4/4$ potential, non-minimal
  coupling to gravity and mixing with the SM Higgs boson}
\label{sec:Model}

The action of the light inflaton model is \cite{Bezrukov:2009yw}
\begin{align}
  S_{X \mathrm{SM}} = &\
    \!\!\int\!\! \sqrt{-g}\, d^4x\left( \L_\mathrm{SM} + \L_{X N}
      + \L_{\text{ext}} + \L_\text{grav} \right)
  , \nonumber \\
  \L_{X N} =          &\ 
     \half \d_\mu X\d^\mu X +\half m_X^2 X^2-\frac{\beta}{4} X^4
     - \lambda \l H^\dagger H - \frac{\alpha}{\lambda} X^2 \r^2
  , \label{4*} \\
  \L_\text{grav} =    &\ -\frac{M_P^2+\xi X^2}{2}R,
\end{align}
where $\L_\text{SM}$ is the SM Lagrangian without the Higgs field
potential, and $\L_{\text{ext}}$ stands for an extension of the SM
capable of explaining all the major phenomenological puzzles: neutrino
oscillations, dark matter phenomena, baryon asymmetry of the Universe
(see discussion in Sec.~\ref{sec:Conclusion}).  As far as we assume
$\alpha,\beta\ll\lambda$, inflation proceeds along the flat direction
of the scalar potential, where the Higgs and inflaton contributions in
the last term of \eqref{4*} cancel.  In the inflaton sector of the
scalar potential the sign of the quadratic term is chosen to lead to
the nonzero vacuum expectation value for the inflaton field at late
stages of the Universe evolution (after reheating).  Then the last
term in \eqref{4*} gives rise to spontaneous breaking of electroweak
(EW) symmetry, and the SM Higgs field gains non-zero vacuum
expectation value as well.  The two scalar excitations above the
vacuum, $h$ and $\chi$, correspond to the SM Higgs boson and inflaton
particles, but the mass basis is slightly rotated as compared to $H$ and
$X$, with the small mixing angle $\theta$.  Four parameters of the
model, $m_X$, $\beta$, $\lambda$, and $\alpha$, determine the Higgs
field vacuum expectation value $v\approx246$\,GeV, the Higgs boson
mass $m_h\approx126$\,GeV, and the inflaton mass
\begin{equation}
  \label{eq:10}
  m_\chi = m_h \sqrt{\frac{\beta}{2\alpha}}
         = \sqrt{\frac{\beta}{\lambda\theta^2}} .
\end{equation}
Thus, at a given value of $\beta$, the only free parameter in the
scalar sector is the mixing coupling $\alpha$ or the inflaton mass
$m_\chi$, which governs the inflaton effective coupling to all the SM
fields via mixing with the Higgs boson characterized by the squared
mixing angle
\begin{equation}
  \label{eq:13}
  \theta^2=\frac{2\beta v^2}{m_\chi^2}=\frac{2\alpha}{\lambda}.
\end{equation}
For zero or very small non-minimal coupling constant ($\xi<10^{-4}$)
the inflationary dynamics is fully determined by the parameter
$\beta$, which is then fixed from the amplitude of primordial density
perturbations.  This amplitude is measured from observations of CMB
anisotropy, and the resulting value, used in
Ref.~\cite{Bezrukov:2009yw}, is $\beta=(1-2)\times10^{-13}$.  In this
case the only free parameter in the model is the inflaton mass
$m_\chi$ (uniquely related to $\alpha$ for given Higgs mass and
$\beta$, see (\ref{eq:13})).  Further, stability of the inflaton
potential with respect to quantum corrections from the SM sector due
to inflaton mixing to the Higgs field places an upper limit on
$\alpha$, while lower limit comes from the requirement of efficient
reheating to the temperature exceeding 100\,GeV (to allow for some
baryogenesis mechanism).  Consequently, using the bounds for $\alpha$,
the inflaton mass was confined in \cite{Bezrukov:2009yw} between
30\,MeV and 2\,GeV.  In Sec.~\ref{sec:Cosmo} we extend those estimates
for larger ($\xi>10^{-4}$) values of non-minimal coupling constant
$\xi$ favored from the combined analysis of present cosmological data
\cite{Hinshaw:2012fq}.

Let us note that generally one more mass parameter in the SM Higgs
sector may be expected (corresponding to the quadratic term for the
Higgs field) and one more mass parameter in the inflaton sector (the
one in front of the cubic term).  The latter is even welcome, as it
helps to get rid of domain walls emerging in the Universe at the
moment when inflaton gains non-zero vacuum expectation value and $Z_2$
symmetry of \eqref{4*} becomes spontaneously broken.  In case the new
dimensionful parameters are of the order of EW scale or below, they do
not change the scale of inflaton mass, but relaxe the relations
between inflation-related and low-energy physics parameters: the value
of inflaton mass would not uniquely determine the inflaton couplings
to other fields in that case.  In what follows we assume those
parameters are much smaller than the EW scale and hence negligible for
our analysis.

\section{Inflation and reheating: prediction of cosmological parameters}
\label{sec:Cosmo}

The inflation in the model happens along the direction
$H^{\dagger}H=\frac{\alpha}{\lambda}X^2$, so that the potential term
in \eqref{4*}, which has a not quite small coupling constant $\lambda$,
vanishes.\footnote{This is true for $\beta\ll1$, which is the case we
  are interested in this article.}  The mass term in \eqref{4*} is
negligible because $m_X$ is significantly below the inflationary
scale.  Thus, the only relevant terms during the inflation are the
quartic potential term and non-minimal coupling with gravity.  This
brings us to the situation thoroughly analyzed in literature (see,
e.g.~\cite{Kaiser:1994vs,Komatsu:1999mt}).  Below in
Sec.~\ref{subsec:indices} we repeat the analysis to arrive at already
known results and obtain some new analytic approximations valid in the
interesting part of the parameter space.

\subsection{Slow roll parameters and spectral indices}
\label{subsec:indices}

To apply the standard slow roll formalism (see
e.g.~\cite{Liddle:1993fq,Gorbunov:2011zzc}) we perform a conformal
transformation of the metric
\begin{equation}
  \label{eq:1}
  g_{\mu\nu}\to  \tilde{g}_{\mu\nu} = \Omega^2\, g_{\mu\nu}\,,
  \qquad
  \Omega^2 = 1+\xi X^2/M_P^2\,,
\end{equation}
In the new variables $\tilde{g}_{\mu\nu}$ (Einstein frame, as opposed
to the original Jordan frame) gravity couples minimally to all matter
fields including inflaton.  Then the inflaton potential gets modified
and reads
\begin{equation}
  \label{eq:2}
  U(X) = \frac{\beta X^4}{4\Omega^4}\,.
\end{equation}
After the conformal transformation the inflaton kinetic term gets
modified too.  In the Einstein frame the canonically normalized
``inflaton field'' ${\cal X}$ is obtained from the solution of
\begin{equation}
  \label{eq:3}
  \frac{d{\cal X}}{d X}
  = \sqrt{\frac{\Omega^2+6\xi^2X^2/M_P^2}{\Omega^4}} .
\end{equation}
Then the potential term for the field ${\cal X}$ is \eqref{eq:2} where
the solution $X=X\l {\cal X}\r$ of eq.~\eqref{eq:3} is substituted.

For the canonically normalized field ${\cal X}$ we can use the
standard formulas for the slow roll parameters during inflation,
number of e-folding and spectral indices.  It has been proven for our
model that prediction for cosmological parameters does not depend on
the choice of frame \cite{Komatsu:1999mt}.  In the Einstein frame we
have a usual large field chaotic inflation with the potential
(\ref{eq:2}), (\ref{eq:3}).  In this case the amplitudes and spectral
indices of scalar and tensor perturbations, prediction of inflation,
are generally determined by the value of the scalar potential and
values of the slow roll parameters.  The slow roll parameters are
\cite{Kaiser:1994vs}
\begin{align}
\label{epsilon}
  \epsilon &
  = \frac{M_P^2}{2}\left(\frac{dU/d{\cal X}}{U}\right)^2
  = \frac{8}{(1+\xi(1+6\xi)X^2/M_P^2)X^2/M_P^2}\,, \\
\label{eta}
  \eta &
  = M_P^2\frac{d^2U/d{\cal X}^2}{U}
  = \frac{4 ( 3+\xi(1+12\xi)X^2/M_P^2-2\xi^2(1+6\xi)X^4/M_P^4 )}
         { (1+\xi(1+6\xi)X^2/M_P^2)^2X^2/M_P^2 }\,.
\end{align}
At the end of inflation $\epsilon=1$ and then from \eqref{epsilon} the
field value $X_{\mathrm{e}}$ is given by \cite{Kaiser:1994vs}
\begin{equation}
  \label{eq:4}
  \frac{X_{\mathrm{e}}^2}{M_P^2} =
    \frac{\sqrt{192\xi^2+32\xi+1}-1}{ 2\xi(1+6\xi) } \,.
\end{equation}
During inflation the inflaton quantum fluctuations become a source of
matter perturbations.  We are interested in the moment during
inflation when perturbations of WMAP pivot scale exit the horizon due
to stretching in almost exponentially expanding Universe.  To estimate
the inflaton field value $X_N$ at the moment, corresponding to $N$
e-foldings before the end of inflation, we should invert the relation
\cite{Kaiser:1994vs}
\begin{multline}
  \label{eq:8}
  N = \int_{X_{\mathrm{e}}}^{X_N}
  \frac{1}{M_P^2}\frac{U}{dU/dX}\left(\frac{d{\cal X}}{dX}\right)^2dX
  = \int_{X_{\mathrm{e}}}^{X_N}
    \frac{X((6\xi^2+\xi)X^2/M_P^2+1)}{4(1+\xi X^2/M_P^2) M_P^2} dX \\
  =
  \frac{3}{4}\left[
    \left(\xi+\frac{1}{6}\right)(X_N^2/M_P^2-X_{\mathrm{e}}^2/M_P^2)
    - \ln\frac{1+\xi X_N^2/M_P^2}
               {1+\xi X_\mathrm{e}^2/M_P^2}
  \right].
\end{multline}
Solving this transcendental equation for $X^2_N$ analytically is
impossible.  To obtain accurate predictions we solve it numerically
and present the results in Fig.~\ref{cosmo-parameters}; below we
describe also the useful analytical approximations.

Now we have all formulas to match parameters of scalar and tensor
perturbation spectra probed in cosmological experiments.  We start
with the perturbation amplitudes.  One adopts the WMAP normalization
of matter perturbation power spectrum \cite{Hinshaw:2012fq} to fix the
ratio
\begin{equation}
  \label{WMAPnorm}
  U/\epsilon=24\pi^2\Delta_\mathcal{R}^2M_P^4\simeq(0.0276\times M_P)^4
  ,
\end{equation}
that in our model \eqref{eq:2}, \eqref{epsilon} is 
\begin{equation}
  \label{eq:7}
  \frac{U}{\epsilon} = 
  \frac{\beta(\xi(6\xi+1)X_N^2/M_P^2+1)X_N^6/M_P^6} 
       {32(1+\xi X_N^2/M_P^2)^2}
\end{equation}
and allows to determine the value of $\beta$ at given $\xi$ and $N$ as
\begin{equation}
  \label{eq:9}
  \beta=24\pi^2\Delta_\mathcal{R}^2 
  \frac{32(1+\xi X_N^2/M_P^2)^2}
       {(\xi(6\xi+1) X_N^2/M_P^2+1)X_N^6/M_P^6}, 
\end{equation}
For numerically solved \eqref{eq:8} at relevant $N=60$ (see below
Sec.~\ref{subsec:reheating}), we present $\beta$ as function of $\xi$
in Fig.~\ref{cosmo-parameters}, lower right panel.

Let us proceed with tensor-to-scalar ratio $r$ and spectral indices of
scalar and tensor perturbations $(n_s-1)$ and $n_T$, respectively. To
the leading order in slow roll parameters \eqref{epsilon},~\eqref{eta}
the general relations are linear (see
e.g.~\cite{Liddle:1993fq,Gorbunov:2011zzc}):
\begin{equation}
\label{slow-roll-predictions}
  r=16\epsilon ,\qquad
  n_s-1=2\eta-6\epsilon ,\qquad
  n_T=-2\epsilon .  
\end{equation}
They must be evaluated at the horizon crossing, when $X=X_N$, see
eq.~\eqref{eq:8}.

Two opposite cases are easy to investigate: models with small and
large values of $\xi$.  Indeed, in the limit $\xi\to 0$ we are back to
the chaotic inflation with quartic scalar potential
\cite{Linde:1983gd}.  Then from eqs.~\eqref{eq:4},~\eqref{eq:8}
$X_N^2/M_P^2\to8(N+1)$ and further from
eqs.~\eqref{epsilon},~\eqref{eta}:
\begin{equation}
  \label{slow-roll-zero}
  \epsilon=\frac{1}{N+1} ,\qquad \eta=\frac{3}{2(N+1)} .
\end{equation}
Plugging these results into eq.~\eqref{slow-roll-predictions} one
arrives at
\begin{equation}
  \label{zero-xi}
  r=\frac{16}{N+1} ,\qquad
  n_s-1=-\frac{3}{N+1} ,\qquad
  n_T=-\frac{2}{N+1} . 
\end{equation}
Eq.~\eqref{eq:9} gives for the quartic self-coupling 
\begin{equation}
  \label{zero-beta}
  \beta=\frac{3\pi^2\,\Delta_\mathcal{R}^2}{2\,(N+1)^3},
\end{equation}
which equals $1.5\times 10^{-13}$ for the relevant value N=60 (see
Sec.~\ref{subsec:reheating}).  It was adopted in
Ref.~\cite{Bezrukov:2009yw} as a reference value for all numerical
estimates.

In the opposite limit $\xi\to \infty$ one obtains from
eqs.~\eqref{eq:4},\eqref{eq:8} at large $N$ (i.e.\ neglecting
logarithmic correction in \eqref{eq:8}): $\xi\,X_N^2/M_P^2\to 4N/3$
and further from eqs.~\eqref{epsilon},\eqref{eta}:
\begin{equation}
  \label{slow-roll-infinity}
  \epsilon=\frac{3}{4N^2} ,\qquad \eta=-\frac{1}{N} .
\end{equation}
Plugging these results into eq.~\eqref{slow-roll-predictions} one
arrives at \cite{Salopek:1992zg,Komatsu:1999mt}
\begin{equation}
  \label{infinite-xi}
  r=\frac{12}{N^2} ,\qquad n_s-1=-\frac{2}{N} ,\qquad
  n_T=-\frac{3}{2 N^2} .
\end{equation}
Self-coupling depends on $\xi$ as follows from eq.~\eqref{eq:9}: 
\begin{equation}
  \label{infinite-beta}
  \beta=\frac{72\pi^2\,\Delta_\mathcal{R}^2}{N^2}\,\xi^2 ,
\end{equation} 
that is $\beta\approx4.5\times10^{-10}\xi^2$ for relevant $N=60$.
Thus even $\beta\sim 1$ is allowed at sufficiently large $\xi$, and
hence the SM Higgs boson non-minimally coupled to gravity may play a
role of inflaton \cite{Bezrukov:2007ep}.  Note that in the limit of
large $\xi$ formulas \eqref{infinite-xi} coincide with those obtained
within $R^2$-inflation
\cite{Starobinsky:1980te,Mukhanov:1981xt,Starobinsky:1983zz}; however,
numerical predictions generically differ because of different
reheating temperature resulting in different values of $N$ in
different models, see discussion in Sec.~\ref{subsec:reheating}.  In
particular, the numerical predictions of $R^2$-inflation
\cite{Mukhanov:1981xt,Starobinsky:1983zz,Gorbunov:2010bn} and
inflation driven by the Higgs boson \cite{Bezrukov:2008ut} differ
\cite{Bezrukov:2011gp,Gorbunov:2012ns}.

To find a reasonably good analytical approximation for the
intermediate case of small but finite value of $\xi$ we utilize the
following relation
\begin{equation}
  \label{interpolating-field}
  \l 1+6\xi\r X_N^2/M_P^2= 8\l N+1\r,
\end{equation}
that follows from eqs.~\eqref{eq:4},~\eqref{eq:8} when linear in $\xi$
corrections are included.  Quite remarkably, though formally obtained
\emph{at small $\xi$,} eq.~\eqref{interpolating-field} interpolates
smoothly between ``$\xi\to0$'' and ``$\xi\to\infty$, large $N$''
regimes considered above, with only $\ln N/N\simeq5\%$ relative error
in $X_N^2$ in the large $\xi$ limit.  Putting
\eqref{interpolating-field} into \eqref{epsilon},~\eqref{eta}, and
\eqref{slow-roll-predictions} one obtains formulas for
tensor-to-scalar ratio $r$ and spectral indices, interpolating between
\eqref{zero-xi} and \eqref{infinite-xi}, cf.~\cite{Anisimov:2008qs}:
\begin{align}
  \label{r-interpolating}
  r     &= \frac{16(1+6\xi)}{(N+1)(1+8(N+1)\xi)} ,\\
  \label{ns-interpolating}
  n_s-1 &= -\frac{3(1+6\xi)+8(N+1)(5+24\xi)\xi+128(N+1)^2\xi^2} 
                 {(N+1)(1+8(N+1)\xi)^2} ,\\
  n_T   &= -\frac{2(1+6\xi)}{(N+1)(1+8(N+1)\xi)} .
  \label{nt-interpolating}
\end{align} 
Similarly one obtains from \eqref{interpolating-field} and
\eqref{eq:9} the formula for inflaton quartic coupling,
\begin{equation}
  \label{beta-interpolating}
  \beta=\frac{3\pi^2\,\Delta_\mathcal{R}^2}{2} 
        \frac{(1+6\xi)(1+6\xi+8(N+1)\xi)}{(1+8(N+1)\xi)(N+1)^3},
\end{equation}
interpolating between \eqref{zero-beta} and \eqref{infinite-beta}.

The approximations \eqref{r-interpolating}--\eqref{beta-interpolating}
are functions of $\xi$ and $N$.  In our model with natural reheating
through the Higgs boson production we obtain $N=60$, see
Sec.~\ref{subsec:reheating}.  In Fig.~\ref{cosmo-parameters},
\FIGURE{
  \includegraphics[width=0.5\textwidth]{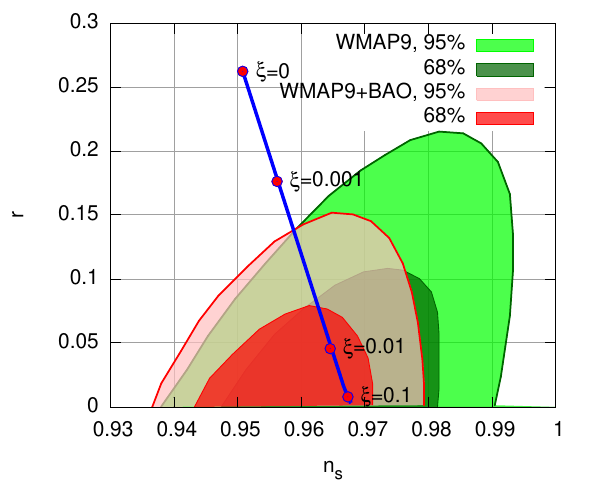}%
  \includegraphics[width=0.5\textwidth]{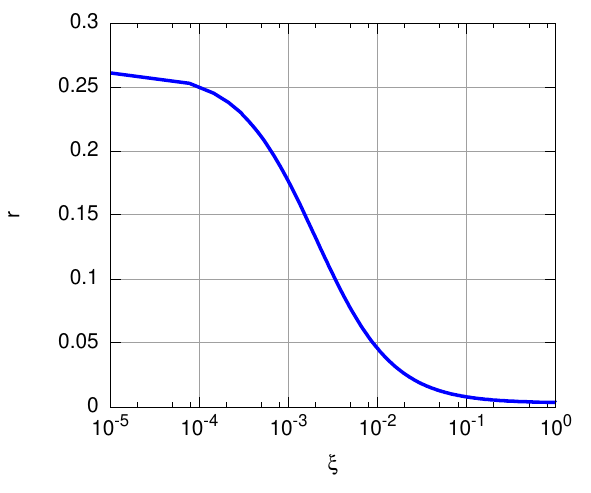}\\
  \includegraphics[width=0.5\textwidth]{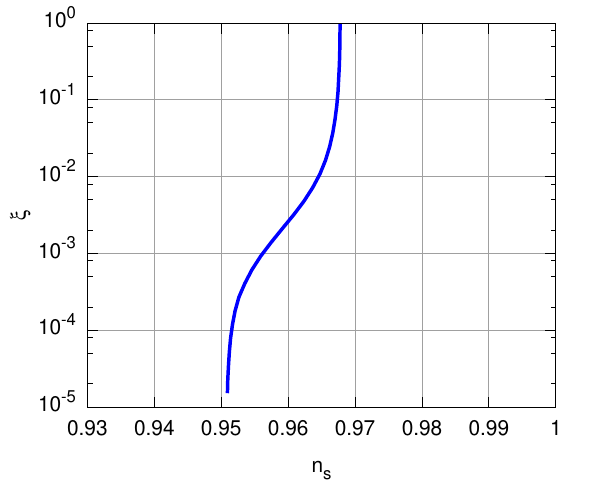}%
  \includegraphics[width=0.5\textwidth]{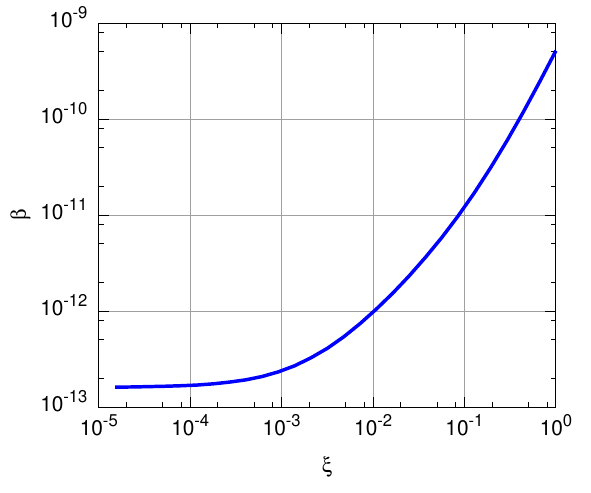}
  \caption{The cosmological parameters for $0\leq\xi<1$; number of
    e-foldings is set to $N=60$ (see discussion in
    Sec.~\protect\ref{subsec:reheating}).  It is clearly seen that
    already at $\xi=0.1$ the model reaches the asymptote of ``large''
    non-minimal coupling.  The shaded regions on the $(r,n_s)$ plot
    are outlined by $1\sigma$- and $2\sigma$-bounds from WMAP9 data
    (green) and WMAP9+BAO data (red)~\cite{Hinshaw:2012fq}.  The
    interval $\xi<2\times 10^{-3}$ is disfavored from cosmology.}
  \label{cosmo-parameters}
}
we present $n_s-1$, $r$ and $\beta$ as \emph{numerical solutions.}
The approximate formulas \eqref{ns-interpolating},
\eqref{r-interpolating}, and \eqref{beta-interpolating}, have relative
errors of not more than 8\%, 4\%, and 8\%, respectively, in the entire
interval of $\xi$ (and are much more precise at low $\xi$).  From
Fig.~\ref{cosmo-parameters} one concludes that the cosmological
parameters effectively reach asymptotes of zero and infinite $\xi$ at
$\xi\lesssim 10^{-4}$ and $\xi\gtrsim 0.1$, respectively.

Note finally, that the analysis in this section can be spoiled by the
radiative corrections from the interaction with the Higgs boson, which
gives the contribution of the order $\alpha^2$ to the inflaton quartic
coupling.  We will somewhat arbitrary use the bound of
$\alpha^2<0.1\beta$ for the radiative corrections to the inflaton.

\subsection{Horizon crossing and reheating}
\label{subsec:reheating}

Reheating in the model happens via the SM Higgs boson production
through inflaton-to-Higgs coupling in \eqref{4*}.  Usually the
inflationary predictions are plotted depending on the number of
e-foldings $N$ that happened after inflaton perturbations of a
specific observable scale exit the horizon.  This number is commonly
estimated as $N=50\text{--}60$.  Let us stress, that in the model of
this article the dynamics of the Universe expansion is known for the
whole time after inflation, including reheating.  Specifically, if the
non-minimal coupling $\xi$ is not too large, post-inflationary
dynamics is dominated by the quartic scalar potential.  This means
that the Universe expands as at the radiation dominated stage.  The
actual reheating moment, though happening quite
late~\cite{Anisimov:2008qs}, leads only to redistribution of energy
between different relativistic degrees of freedom, and does not
influence the rate of the Universe expansion.

Let us derive the relation between the number of e-foldings $N$ and
model parameters.  We are interested in the moment when the
perturbation modes of conformal momentum $k$, corresponding to the
WMAP pivot scale $k/a_0=0.002\text{\,Mpc}^{-1}$ \cite{Hinshaw:2012fq},
exit the horizon at inflationary stage.  Hereafter $a_0$ is the
present scale factor, ``$e$'' and asterisk in subscript refer to the
values at the end of inflation and at the horizon crossing of the
modes defined above; so $N=\log(a_e/a_*)$ by definition.  At horizon
crossing the Hubble parameter ${\cal H}$ obeys the equation
\[
  {\cal H}_*=\frac{k}{a_*}\equiv\frac{k}{a_e}\,\e^N.
\]
Then  
\begin{equation}
  \label{e-foldings}
  N=\log \frac{a_0{\cal H}_*}{k}+\log\frac{a_r}{a_0}+\log\frac{a_e}{a_r},
\end{equation}
where ``$r$'' in subscript refers to the values at reheating.  At the
hot stage entropy in comoving volume is conserved, hence
\begin{equation}
  \label{scale-factors-at-reheating}
  \frac{a_r}{a_0}=\frac{T_0}{T_r}\,\frac{g_0^{1/3}}{g_r^{1/3}},
\end{equation}
where for the effective number of the SM degrees of freedom at present
time and at reheating one has $g_0=43/11$ and $g_r=106.75$,
respectively.  Since between inflation and reheating the Universe in
our model expands as it would be at radiation domination, we have
\begin{equation}
  \label{scale-factors-after-inflation}
  \frac{a_e}{a_r}=\frac{T_r}{U_e^{1/4}}\,\frac{g_r^{1/4}\pi^{1/2}}{30^{1/4}}.
\end{equation}
Plugging eqs.~\eqref{scale-factors-at-reheating},
\eqref{scale-factors-after-inflation} into \eqref{e-foldings} and
substituting the Hubble parameter from the Friedman equation we obtain
\begin{equation}
  \label{e-foldings-final}
  N=\log
  \frac{T_0\,g_0^{1/3}}{k/a_0}+\frac{1}{4}\, 
  \log\frac{\pi^2}{270\,g_r^{1/3}} + 
  \log\frac{U_*^{1/2}}{M_P\,U_e^{1/4}},
\end{equation}
where $T_0$ is CMB temperature at present.  Note that dependence of
$N$ on the reheating temperature enters only through the number of degrees
of freedom at reheating $g_r$.  It is due to the fact that at hot
stage the expansion is not isothermal but an adiabatic process, so the
cosmological observables depends on the effective numbers of degrees
of freedom in plasma at the onset of the hot stage $g_r$ and at
present $g_0$.
 
Numerically, for WMAP pivot scale $k/a_0=0.002\text{\,Mpc}^{-1}$
\cite{Hinshaw:2012fq}, one gets
\begin{equation}
  \label{e-foldings-numeric}
  N=64.3+ 
  \log\frac{U_*^{1/4}/M_P}{\epsilon_*^{1/4}} 
  + \log\frac{U_*^{1/4}\,\epsilon_*^{1/4}}{U_e^{1/4}}.
\end{equation}
The first logarithmic term in \eqref{e-foldings-numeric} follows from
eq.~\eqref{WMAPnorm},
$\log\frac{U_*^{1/4}/M_P}{\epsilon_*^{1/4}}=-3.6$.  Substituting
eqs.~\eqref{epsilon}, \eqref{eq:2}, \eqref{eq:4} into the last
logarithmic term in \eqref{e-foldings-numeric}, one arrives at
\begin{equation}
  \label{e-foldings-exact}
  N=60.7
  +\frac{1}{2}\,\log\frac{(1+16\,\xi+\sqrt{1+32\,(1+6\,\xi)\,\xi}) 
    \,X_N/M_P} 
  {8\,\sqrt{2}\,\sqrt{1+(1+6\,\xi)\,\xi\,X_N^2/M_P^2}\,(1+\xi\,X_N^2/M_P^2)}.
\end{equation}
We use this formula for the numerical estimates of the number of
e-foldings in a model with non-zero finit $\xi$.
One gets $N=61.5\text{--}58.5$ for $\xi$ changing from zero to infinity.

Let us analyze briefly the reheating itself.  After the inflation the
action for the scalar sector in the Einstein frame is
\[
  S = \int d^4x\left\{
    \frac{M_P^2}{2}R+\frac{\dm H^\dagger\dm H}{\Omega^2}+
    \frac{\Omega^2+6\xi^2X^2/M_P^2}{\Omega^4}\frac{\dm X\dm X}{2}
    -\frac{V(H,X)}{\Omega^4}
  \right\}.
\]
In the reheating regime $X<X_e$ for $\xi<1$ we can neglect the
coefficient in front of the kinetic term for $X$: the Universe is
dominated by the field $X$ oscillating due to the quartic potential.
For zero $\xi$ the reheating due to the potential mixing term
${\alpha}X^2H^{\dagger}H$ was analyzed in detail in
\cite{Anisimov:2008qs}, and leads to the bound
\[
  \alpha>0.7\times10^{-11}.
\]
For $\xi\neq0$ additional contributions
to the scattering process appear suppressed\footnote{Let us note here
  the significant difference from the situation of $\xi\gg1$, which is
  similar to the case of $R^2$ inflation
  \cite{Gorbunov:2010bn,Gorbunov:2012ns}.  For $\xi\gg1$ the kinetic
  term us significantly modified, and the field redefinition for $X$
  is needed during reheating, leading to a much weaker suppression of
  the interactions with the Higgs boson by first power of $M_P$.} by
at least $M_P^2/\xi$.  It is easy to estimate that this contribution
is negligible compared to the quartic mixing term and all the bounds
on $\alpha$ in \cite{Anisimov:2008qs} remain valid.

\section{Phenomenology: searches in particle physics experiments}
\label{sec:Pheno}

We see that models with very small non-minimal coupling
$\xi<2\times10^{-3}$ are disfavored from cosmology (upper panels in
Fig.~\ref{cosmo-parameters}) and hence the relevant for low energy
phenomenology quartic self-coupling $\beta$ should obey
$\beta>3\times10^{-13}$ (lower right panel in
Fig.~\ref{cosmo-parameters}).  Accounting for $\xi$-dependence of
$\beta$ in eq.~\eqref{eq:10}, where the SM Higgs boson mass is
$m_h=126$\,GeV, and recalling that parameter $\alpha$ is bounded from
above and below by constraints on quantum corrections and reheating
temperature, respectively, as explained in Sec.~\ref{sec:Model}, we
end up with cosmologically motivated interval of the inflaton mass
$m_\chi$, confined between green and blue shaded regions of
Fig.~\ref{fig:inflaton-all-bounds}
\FIGURE{
  \includegraphics[width=0.5\textwidth]{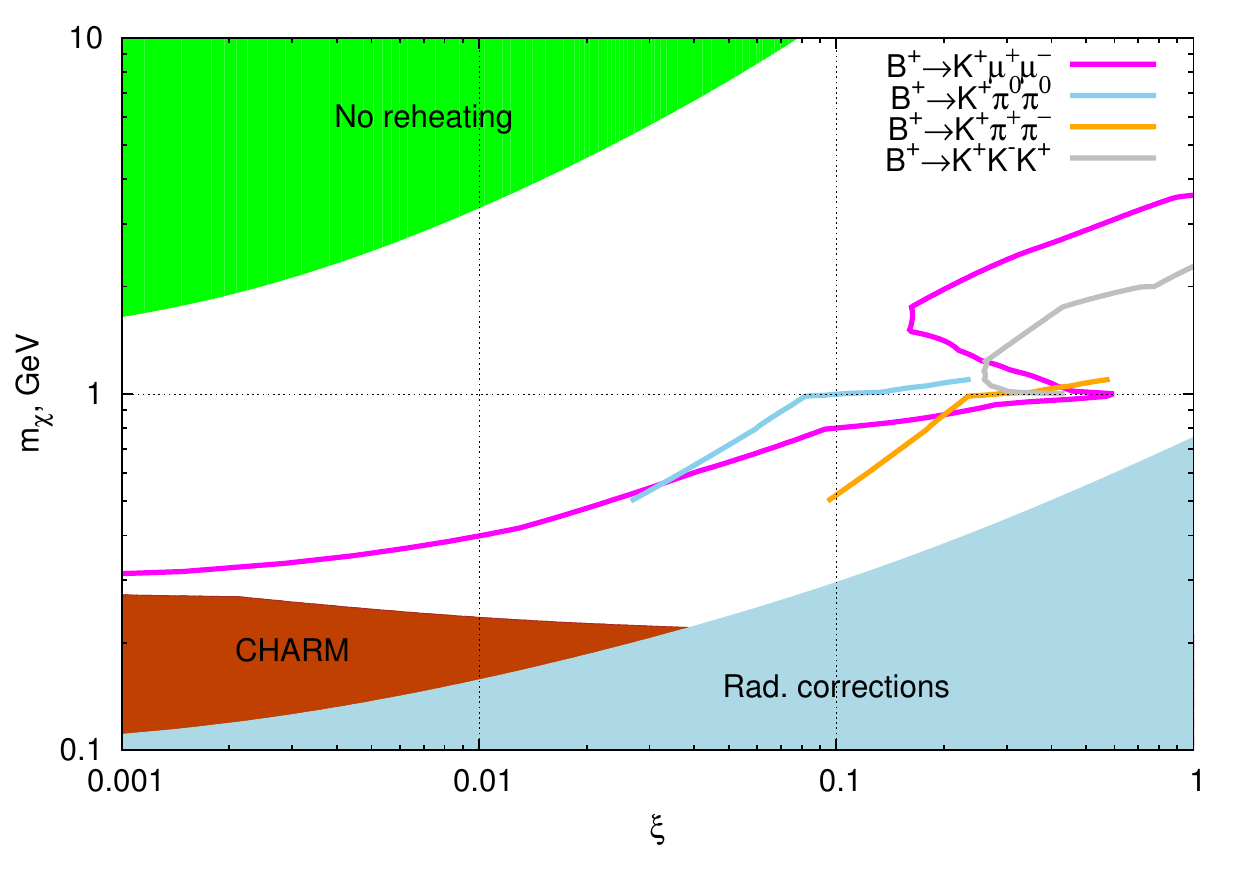}%
  \includegraphics[width=0.5\textwidth]{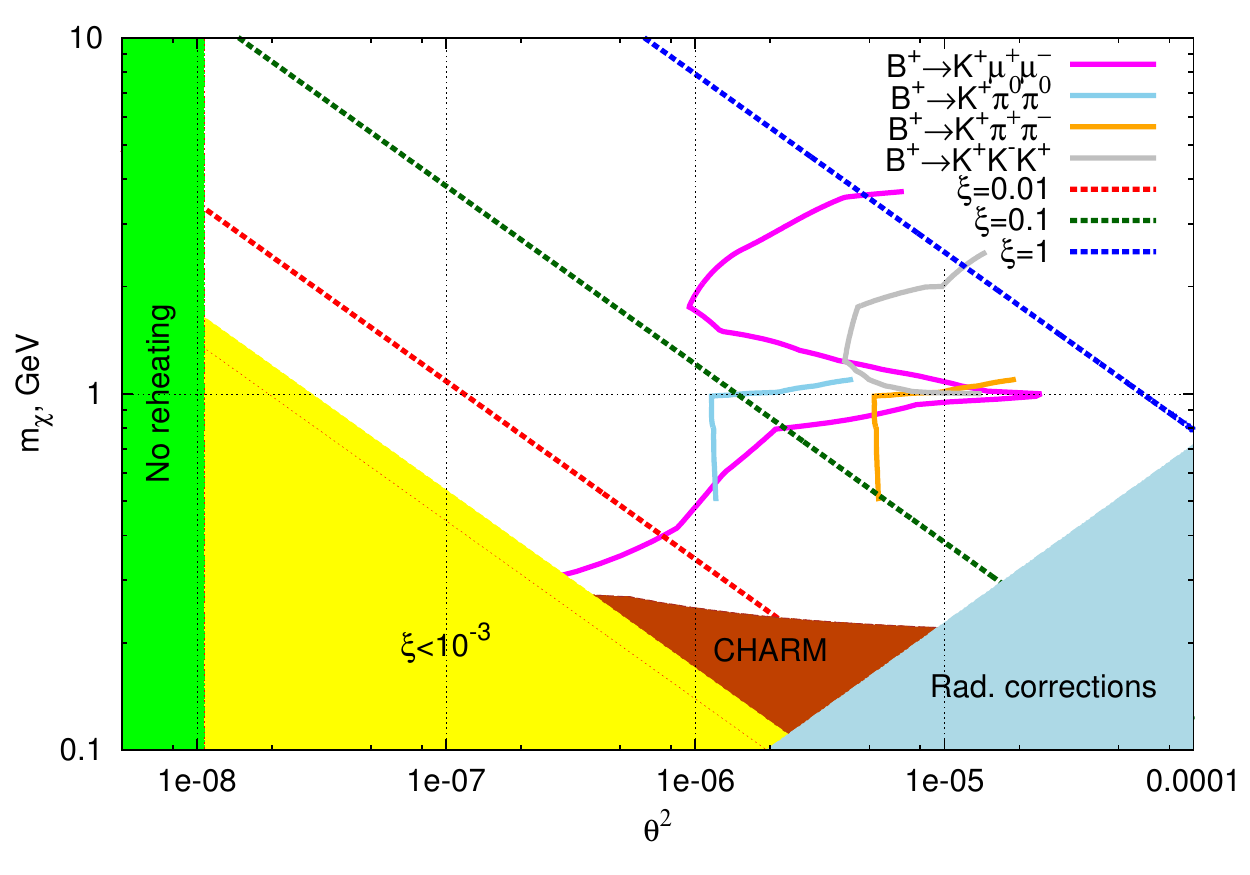}\\
  \includegraphics[width=0.5\textwidth]{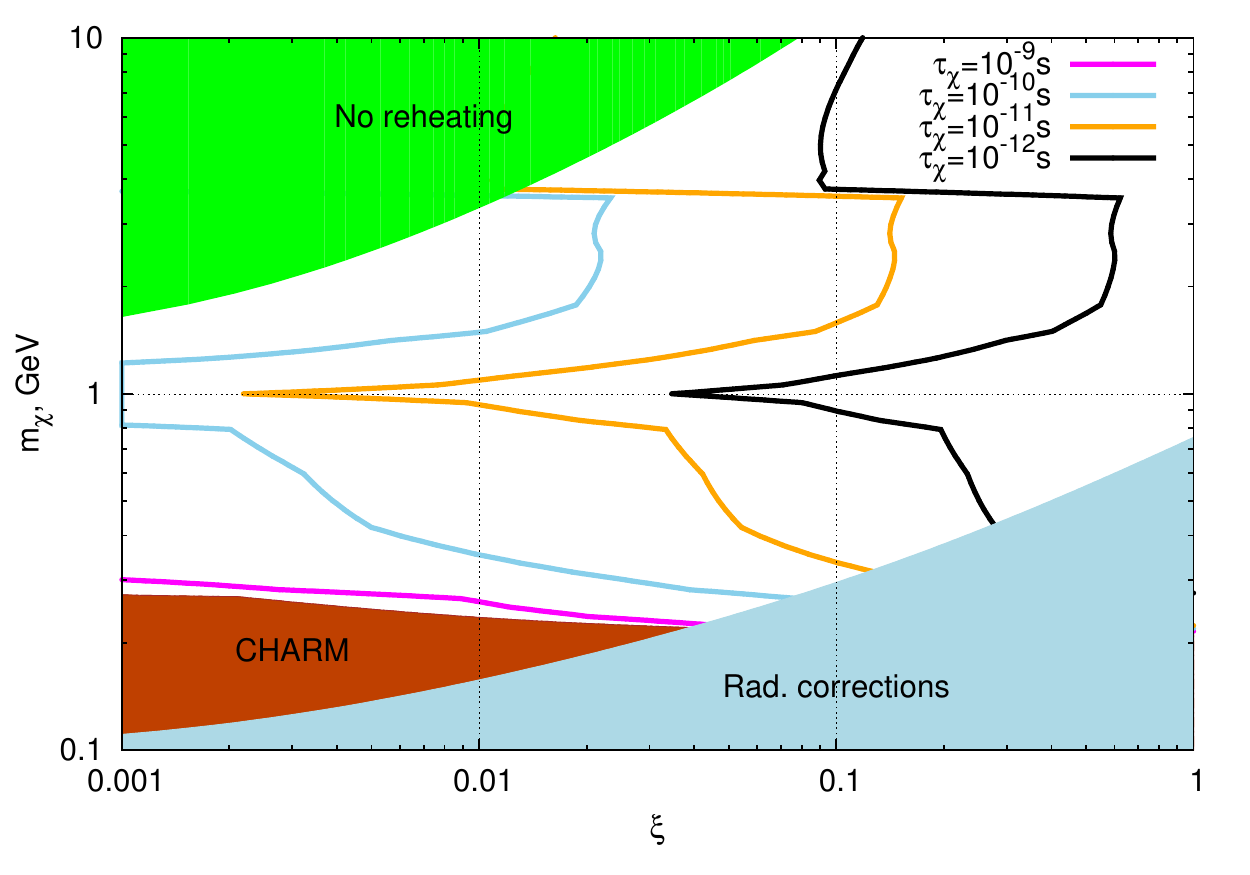}%
  \includegraphics[width=0.5\textwidth]{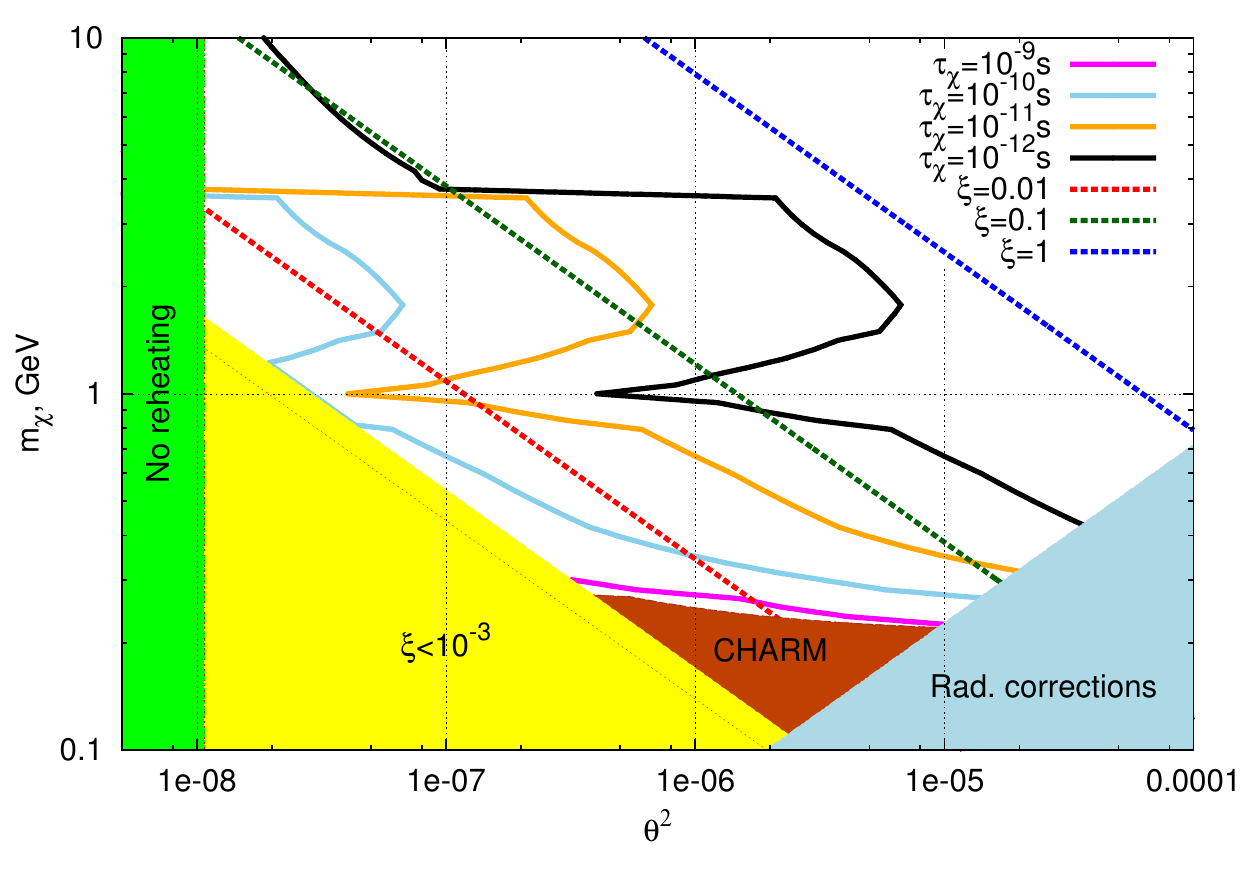}
  \caption{The bounds on the inflaton in the planes $(m_\chi,\xi)$ and
    $(m_\chi,\theta^2)$.  Cosmological constraints are shaded: the
    green region leads to insufficient reheating, the blue region
    gives large radiative corrections to inflaton potential; the
    region $\xi<2\times 10^{-3}$ is disfavored from analysis of CMB
    data presented in Sec.~\protect\ref{sec:Cosmo}.  Other constraints
    are from direct searches in accelerator experiments: the brown
    region is excluded by the CHARM experiment.  On the upper two
    plots other lines are the bounds from $B$ meson decays, with the
    excluded areas to the right of the lines.  On the lower plots the
    lines correspond to inflaton lifetime.  Dashed lines on the right
    plots are isolines of constant $\xi$.}
  \label{fig:inflaton-all-bounds}
}
at $\xi>2\times 10^{-3}$.   

The upper limit on $m_\chi$ at a given $\xi$ comes from requirement of
sufficiently energetic reheating, that develops in the same way as in
the case of $\xi=0$, studied in Ref.~\cite{Anisimov:2008qs}.  It
implies a limit $\alpha>0.7\times 10^{-11}$, which is substituted to
\eqref{eq:10} and with account of $\xi$-dependence of $\beta$ is
plotted in Fig.~\ref{fig:inflaton-all-bounds}. The lower limit on
$m_\chi$ for the case of $\xi=0$ has been obtained in
\cite{Bezrukov:2009yw} from smallness of quantum corrections to the
inflaton potential, guaranteed at
\[
  \alpha^2<0.1\times\beta.
\]
While $\xi<1$, that estimate remains valid. We plug it into
eq.~\eqref{eq:10} and recall $\xi$-dependence of $\beta$ to depict the
upper limit on inflaton mass in Fig.~\ref{fig:inflaton-all-bounds}.

Another set of bounds on the model parameters comes from particle
physics experiments.  Thanks to the inflaton-Higgs mixing
\eqref{eq:13}, inflaton can be produced in the same channels as the SM
Higgs boson would be of the same mass $m_\chi$. The same mixing is
responsible for the inflaton decay to the light SM particles, and its
decay pattern is exactly the same as that of the normal SM Higgs
boson, if its mass would be equal to $m_\chi$. Inflaton decay
branching ratios are calculated in Ref.~\cite{Bezrukov:2009yw} and
here we reproduce them in Fig.~\ref{inflaton-branchings}.
\FIGURE{
  \centerline{
    \includegraphics{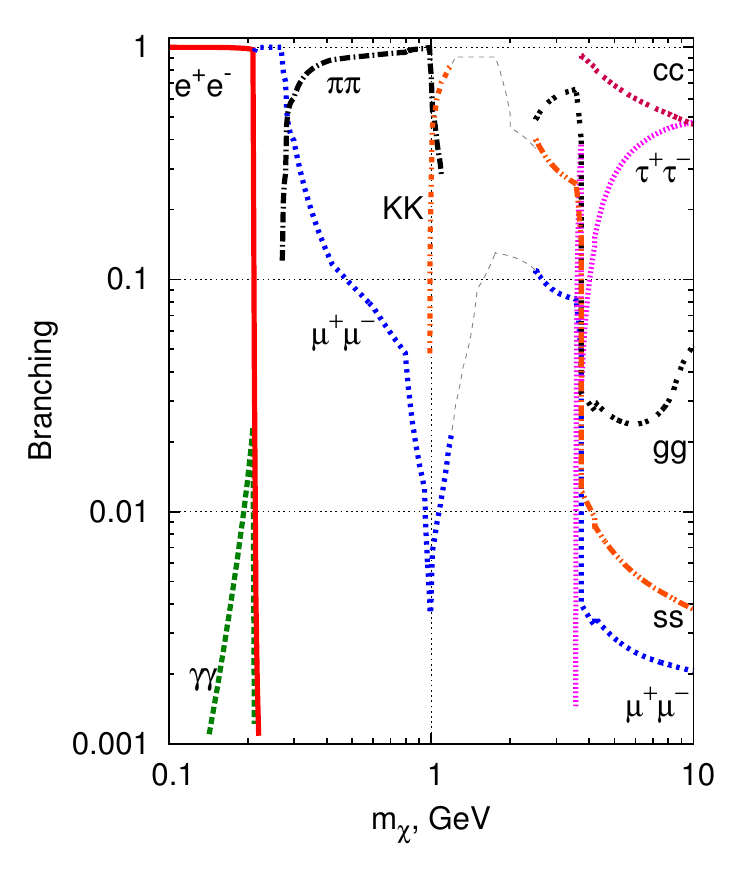}%
    \includegraphics{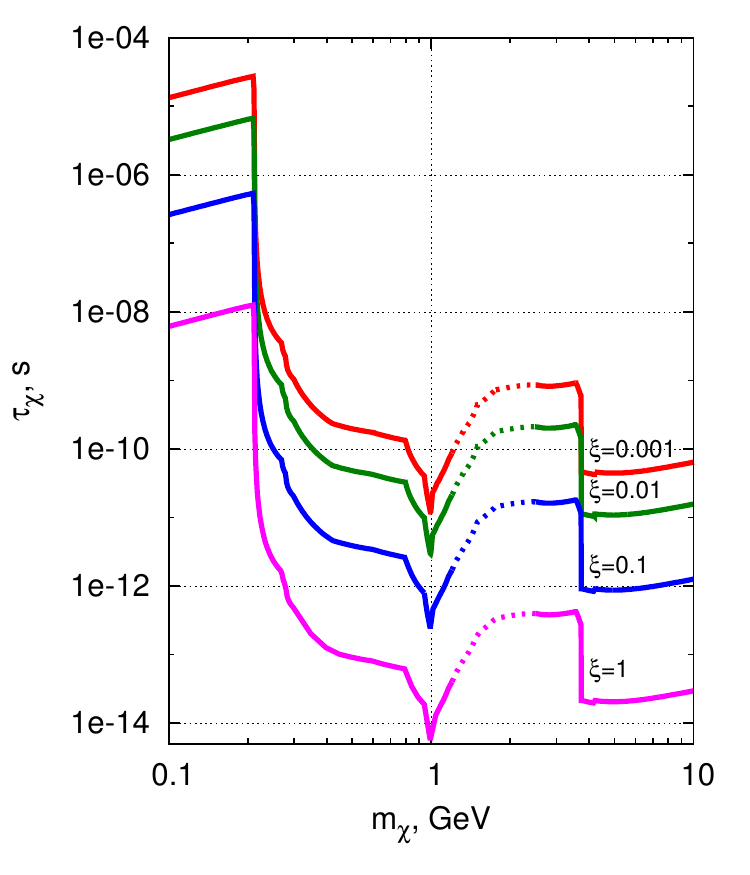}
  }
  \caption{Inflaton decay branching ratios (\emph{left plot}) and
    inflaton lifetime (\emph{right plot}); theoretical predictions for
    $m_\chi\simeq 1-2$\,GeV (thin dashed lines on the \emph{left plot}
    and dotted lines on the \emph{right plot}) suffer from significant
    QCD-uncertainties.}
  \label{inflaton-branchings}
}
Note that in the inflaton mass range 1-2\,GeV the estimates are rather
vague because of QCD-uncertainties with hadronization.  The inflaton
lifetime depends on $\beta$ and hence on $\xi$ given the relation in
Fig.~\ref{cosmo-parameters} (right lower panel).  We have recalculated
inflaton lifetime by replacing $\beta\to\beta(\xi)$ in formulas of
Ref.~\cite{Bezrukov:2009yw}, the result is plotted in
Fig.~\ref{inflaton-branchings} (right panel).

Inflaton-to-Higgs mixing opens a room for direct searches for the
light inflaton, which has been thoroughly studied in
Ref.~\cite{Bezrukov:2009yw}.  Remarkably, only $m_\chi$ and $\beta$
enter inflaton-Higgs mixing \eqref{eq:13}, hence no dependence on the
Higgs boson mass in the inflaton production rate or its decay
branching ratios.  Thus laboratory experiments allow for probing
directly the inflaton quartic coupling responsible for the
inflationary stage in the early Universe.  In~\cite{Bezrukov:2009yw}
we have found as most relevant the analysis of the CHARM bound.  In
case of non-zero $\xi$ it goes along the same lines as in Sec.~6 of
Ref.~\cite{Bezrukov:2009yw}, with the only substitution of
$\beta\to\beta(\xi)$ in the inflaton decay length and meson decay
branching ratios into inflaton.  The resulting bound on $\beta(\xi)$
is read from Fig.~5 of Ref.~\cite{Bezrukov:2009yw} and thus we obtain
the lower limit on the inflaton mass at given $\xi$ depicted in
Fig.~\ref{fig:inflaton-all-bounds}.  One observes in
Fig.~\ref{fig:inflaton-all-bounds}, that with increasing $\xi$ at first
the lower masses become allowed, but quite fast the radiative
corrections bound for the inflation starts to dominate.  Moreover,
the whole inflaton window moves to higher masses (that was in fact
known, see~\cite{Anisimov:2008qs}).

The most promising processes in searches for the inflaton of
$m_\chi<5$\,GeV are $b$-quark decays.  In Ref.~\cite{Bezrukov:2009yw}
we found for the $B$-meson decay rate
\begin{align}
  \label{B-branching}
  \text{Br}(B\to\chi X_s) &\simeq 10^{-6}\times \l 1-
  \frac{m_\chi^2}{m_b^2}\r^2 \l \frac{\beta(\xi)}{1.5\times 10^{-13}}\r 
  \l \frac{300\MeV}{m_\chi}\r^2 \\
 &\simeq 4.8\times10^{-6}\times \l 1-
  \frac{m_\chi^2}{m_b^2}\r^2 \l \frac{\theta^2}{10^{-6}}\r.
  \notag
\end{align}
Here $X_s$ refers to a strange hadron, which most probably turns out
to be a $K$-meson, because the other decay product, $\chi$, is a
scalar.  For a not so small value of $\xi$ the inflaton is short lived
(see Fig.~\ref{inflaton-branchings}) and decays inside the detector
into SM particles,
\[
  \chi \to \mu^+\mu^-\,,\;\pi^+\pi^-\,,\;\pi^0\pi^0 \,,\;K^+K^-\,,\dots,  
\]
thus contributing to the corresponding three-body decay modes
of $B$-meson.  Conservatively, we require that inflaton contributions,
estimated with eq.~\eqref{B-branching} and decay branching ratios in
Fig.~\ref{inflaton-branchings}, are smaller than the
measured branching ratios of corresponding non-resonant three-body
decays of $B$-mesons \cite{Beringer:1900zz}, specifically
\begin{align}
\label{11}
  \text{Br}(B\to K^+\pi^0\pi^0) &< 1.6\times10^{-5},\\
  \text{Br}(B\to K^+\pi^-\pi^+ \text{ nonresonant})
                                &< 1.6\times10^{-5},\\
  \text{Br}(B\to K^+\mu^+\mu^-) &< 4.8\times10^{-7},\\
  \text{Br}(B\to K^+K^-K^+ \text{ nonresonant})
                                &< 2.8\times10^{-5}.
\label{22}
\end{align}
Resulting limits on model parameters are presented in
Fig.~\ref{fig:inflaton-all-bounds}.  One can see, that a wide range of
inflaton masses $300\MeV\lesssim m_\chi\lesssim 5$\,GeV is allowed for
direct investigation at $B$-factories. The branching ratio of the
two-body $B$-meson decay is at the level of $10^{-5}-10^{-8}$ for
$\beta(\xi)$ in Fig.~\ref{cosmo-parameters}.

Note in passing that our limits from $B$-physics are only approximate.
First, they do not apply to the inflaton masses close to the masses of
known hadronic resonances, since limits \eqref{11}--\eqref{22} are
obtained from the analysis of corresponding Dalitz plots with removed
resonances.  Second, the absence of new resonances on the Dalitz plots
(would it be specially studied) put stronger constraints on the light
inflaton model, than what we naively extracted from
\eqref{11}--\eqref{22}.  Third, one should be careful about the bounds
from $B\to K\mu\mu$, because it is not valid for living inflatons
$\tau_\chi\gtrsim10^{-9}$, which can escape the detector (the lower
region near the CHARM bound).

\section{Discussion}
\label{sec:Conclusion}

To summarize, following the new experimental results both in particle
physics \cite{:2012gk,:2012gu} and cosmology
\cite{Story:2012wx,Hinshaw:2012fq,Sievers:2013wk}, we have extended
our previous analysis \cite{Bezrukov:2009yw} of the large-field
chaotic inflationary model with quartic inflaton potential to the case
of inflaton non-minimally coupled to gravity.  We concluded, that the
range of cosmologically acceptable masses for the inflaton is wider,
while the coupling between the inflaton and the SM sector can be
stronger, than suggested \cite{Bezrukov:2009yw} for the minimally
coupled case.  The model can be further tested with new cosmological
data, especially with further results on the tensor perturbations of
the CMB.

The best place to probe the model remains the same: studies of
two-body $B$-meson decays to kaons and inflatons, where the inflaton
decays further into electron, muon, pion or kaon pair.  However with
the increase of the non-minimal coupling $\xi$ the inflaton lifetime
decreases.  In a smaller part of parameter space the inflaton lifetime
is long enough to cover a macroscopic distance from the $B$-meson
decay position, which leads to the following signatures of the
inflaton event: either missing energy (if the inflaton escapes from
the detector) or significantly displaced vertex of the inflaton decay.
Therefore, searches for process $B\to K+\text{nothing}$ with
\emph{two-body kinematics} and $B\to K+\chi$ with subsequent decay of
$\chi$ to a pair of SM light particles at a macroscopic distance from
the $B$-meson decay are necessary to probe the model.  In a wider
region of parameter space the inflaton decays close to the $B$-meson
decay point, thus effectively contributing to the three-body $B$-meson
decays.  This gives a possibility to probe the model by investigating
neutral two-body resonance-like events ($\mu^+\mu^-$, $\pi^+\pi^-$,
etc.) among three-body final states of $B$-meson decays.  The typical
distance to the inflaton decay vertex, distinguishing these two
options, can be deduced from the lifetime of the inflaton, see
Figs.~\ref{fig:inflaton-all-bounds},~\ref{inflaton-branchings}.

Independently (and also if $B$-meson decays are kinematically
forbidden) the model can be probed to some extent in a fixed-target
experiment.  The inflaton can be produced by protons on target either
in meson decays or through the gluon fusion similar to the SM Higgs
boson, would it be of the same mass.  This option has been studied in
detail in Ref.~\cite{Bezrukov:2009yw} for the case of the inflaton
minimally-coupled to gravity.  The estimates of the inflaton
production cross sections for a realistic set of beams are given there
in Fig.~4.  For the non-minimally coupled inflaton those estimates
have to be multiplied by a factor $\beta(\xi)/(1.5\times10^{-13})$,
which scales from 1 to $3\times 10^3$ when $\xi$ grows from 0 to 1,
see Fig.~\ref{cosmo-parameters}.

Note also, that the model depends on the positivity of the SM Higgs
boson self-coupling for the whole range of energy scales, relevant for
inflation (or, to be more precise, for the scales of the order
$\sim\sqrt{\alpha/\lambda}M_P$). It implies a lower limit on the Higgs
boson mass.  This limit is slightly weaker than the bounds from the
positivity of the Higgs self-coupling up to the Planck scale (see
\cite{Bezrukov:2012sa,Degrassi:2012ry,Alekhin:2012py}), and is
compatible with the observations at LHC within presently achieved
precision for the Higgs and top-quark masses.

Finally, the inflationary model under discussion can be further
modified (completed) to solve other phenomenological problems of the
SM (neutrino oscillations, dark matter phenomena, baryon asymmetry of
the Universe) in such a way that the inflationary dynamics and the
low-energy phenomenology of the light inflaton remains unchanged.
The working example is an extension of the SM with three Majorana
sterile neutrinos, which can explain neutrino oscillations via type-I
seesaw mechanism, the baryon asymmetry via resonant leptogenesis in
the early Universe \cite{Pilaftsis:1997jf,Asaka:2005pn} and the
lightest of the sterile neutrinos can serve as dark matter. With
sterile neutrinos in GeV mass range the model is known as $\nu$MSM
\cite{Asaka:2005an,Asaka:2005pn} and thoroughly studied
\cite{Boyarsky:2009ix}, though both seesaw mechanism and resonant
leptogenesis allow for (much) heavier neutrinos as well.  Note, that
within our logic of only one energy scale in the whole model
advertised in Introduction and Sec.~\ref{sec:Model}, the natural mass
scale for these neutrinos should not exceed the Electroweak scale.

Moreover, the inflaton being singlet with respect to the SM gauge
group may have Yukawa coupling to the sterile neutrino and hence
contributes to their production in the early Universe and their masses
after getting non-zero vacuum expectation value. In this way one can
avoid the explicit mass term for the sterile neutrino and deal with
the explicitly one-scale model of particle physics.  This particular
setup has been considered in Appendix A of Ref.~\cite{Bezrukov:2009yw}
and all formulas presented there can be fully transferred to the case
of the non-minimally coupled to gravity inflaton, since all of them
explicitly contain the factor $\beta=\beta(\xi)$ and the inflaton
mass.  Most importantly, in the non-minimally coupled case as well,
the inflaton may contribute to the dark matter production (the
lightest sterile neutrino) thus eluding the strong fine-tuning
\cite{Canetti:2012kh,Roy:2010xq} required in the simplest version of
$\nu$MSM to generate the dark matter. Remarkably, the Yukawa
interactions of inflaton with sterile neutrinos is bounded from the
requirement of small radiative corrections to the inflationary
potential, and \emph{can not give rise to very large masses}
\cite{Bezrukov:2009yw,Shaposhnikov:2006xi} consistently with the logic
adopted.

However, explicit mass parameters in the sterile neutrino part of the
action are generally allowed.  The only theoretical bound on them
comes from the smallness of generated by the seesaw couplings one-loop
finite corrections to the SM Higgs boson mass.  The bound allows the
sterile neutrino masses to be well above the EW scale.  In this setup
the mechanism of generation of the dark matter sterile neutrino in the
inflaton decays still works, as well as the resonant leptogenesis
\cite{Pilaftsis:1997jf}. We do not study this setup further as one
beyond our logical framework.

\acknowledgments
 
We thank P.~Pakhlov for a valuable discussion on B-factories. 
The work of D.G. is supported in part by the grant of the
President of the Russian Federation NS-5590.2012.2, by RFBR grants 
11-02-01528a, 13-02-01127a, and by MSE under contract \#8412.

\bibliographystyle{JCAP-hyper}
\bibliography{refs}

\providecommand{\href}[2]{#2}\begingroup\raggedright\begin{thebibliography}{10}

\bibitem{Bezrukov:2009yw}
F.~Bezrukov and D.~Gorbunov, {\it {Light inflaton Hunter's Guide}},
  \href{http://dx.doi.org/10.1007/JHEP05(2010)010}{{\em JHEP} {\bf 1005} (2010)
   010}, [\href{http://xxx.lanl.gov/abs/0912.0390}{{\tt arXiv:0912.0390}}],
  \href{http://www.slac.stanford.edu/spires/find/hep/www?rawcmd=eprint+0912.0390}{[SPIRES]}.

\bibitem{Starobinsky:1980te}
A.~A. Starobinsky, {\it {A New Type of Isotropic Cosmological Models Without
  Singularity}},  \href{http://dx.doi.org/10.1016/0370-2693(80)90670-X}{{\em
  Phys.Lett.} {\bf B91} (1980)  99--102},
  \href{http://www.slac.stanford.edu/spires/find/hep/www?rawcmd=j+PHLTA,B91,99}{[SPIRES]}.

\bibitem{Mukhanov:1981xt}
V.~F. Mukhanov and G.~Chibisov, {\it {Quantum Fluctuation and Nonsingular
  Universe. (In Russian)}},  {\em JETP Lett.} {\bf 33} (1981)  532--535,
  \href{http://www.slac.stanford.edu/spires/find/hep/www?rawcmd=j+JTPLA,33,532}{[SPIRES]}.

\bibitem{Guth:1980zm}
A.~H. Guth, {\it The inflationary universe: A possible solution to the horizon
  and flatness problems},  {\em Phys. Rev.} {\bf D23} (1981)  347--356,
  \href{http://www.slac.stanford.edu/spires/find/hep/www?rawcmd=j+PHRVA,D23,347}{[SPIRES]}.

\bibitem{Linde:1981mu}
A.~D. Linde, {\it A new inflationary universe scenario: A possible solution of
  the horizon, flatness, homogeneity, isotropy and primordial monopole
  problems},  {\em Phys. Lett.} {\bf B108} (1982)  389--393,
  \href{http://www.slac.stanford.edu/spires/find/hep/www?rawcmd=j+PHLTA,B108,389}{[SPIRES]}.

\bibitem{Albrecht:1982wi}
A.~Albrecht and P.~J. Steinhardt, {\it Cosmology for grand unified theories
  with radiatively induced symmetry breaking},  {\em Phys. Rev. Lett.} {\bf 48}
  (1982)  1220--1223,
  \href{http://www.slac.stanford.edu/spires/find/hep/www?rawcmd=j+PRLTA,48,1220}{[SPIRES]}.

\bibitem{Shaposhnikov:2006xi}
M.~Shaposhnikov and I.~Tkachev, {\it The {$\nu$MSM}, inflation, and dark
  matter},  \href{http://dx.doi.org/10.1016/j.physletb.2006.06.063}{{\em Phys.
  Lett.} {\bf B639} (2006)  414--417},
  [\href{http://xxx.lanl.gov/abs/hep-ph/0604236}{{\tt hep-ph/0604236}}],
  \href{http://www.slac.stanford.edu/spires/find/hep/www?rawcmd=eprint+hep-ph\%2F0604236}{[SPRIES]}.

\bibitem{:2012gk}
{\bf ATLAS Collaboration} Collaboration, G.~Aad {\em et.~al.}, {\it
  {Observation of a new particle in the search for the Standard Model Higgs
  boson with the ATLAS detector at the LHC}},
  \href{http://dx.doi.org/10.1016/j.physletb.2012.08.020}{{\em Phys.Lett.} {\bf
  B716} (2012)  1--29}, [\href{http://xxx.lanl.gov/abs/1207.7214}{{\tt
  arXiv:1207.7214}}],
  \href{http://www.slac.stanford.edu/spires/find/hep/www?rawcmd=eprint+1207.7214}{[SPIRES]}.

\bibitem{:2012gu}
{\bf CMS Collaboration} Collaboration, S.~Chatrchyan {\em et.~al.}, {\it
  {Observation of a new boson at a mass of 125 GeV with the CMS experiment at
  the LHC}},  \href{http://dx.doi.org/10.1016/j.physletb.2012.08.021}{{\em
  Phys.Lett.} {\bf B716} (2012)  30--61},
  [\href{http://xxx.lanl.gov/abs/1207.7235}{{\tt arXiv:1207.7235}}],
  \href{http://www.slac.stanford.edu/spires/find/hep/www?rawcmd=eprint+1207.7235}{[SPIRES]}.

\bibitem{Story:2012wx}
K.~Story, C.~Reichardt, Z.~Hou, R.~Keisler, K.~Aird, {\em et.~al.}, {\it {A
  Measurement of the Cosmic Microwave Background Damping Tail from the
  2500-square-degree SPT-SZ survey}},
  \href{http://xxx.lanl.gov/abs/1210.7231}{{\tt arXiv:1210.7231}},
  \href{http://www.slac.stanford.edu/spires/find/hep/www?rawcmd=eprint+1210.7231}{[SPIRES]}.

\bibitem{Sievers:2013wk}
J.~L. Sievers, R.~A. Hlozek, M.~R. Nolta, V.~Acquaviva, G.~E. Addison, {\em
  et.~al.}, {\it {The Atacama Cosmology Telescope: Cosmological parameters from
  three seasons of data}},  \href{http://xxx.lanl.gov/abs/1301.0824}{{\tt
  arXiv:1301.0824}},
  \href{http://www.slac.stanford.edu/spires/find/hep/www?rawcmd=eprint+1301.0824}{[SPIRES]}.

\bibitem{Hinshaw:2012fq}
G.~Hinshaw, D.~Larson, E.~Komatsu, D.~Spergel, C.~Bennett, {\em et.~al.}, {\it
  {Nine-Year Wilkinson Microwave Anisotropy Probe (WMAP) Observations:
  Cosmological Parameter Results}},
  \href{http://xxx.lanl.gov/abs/1212.5226}{{\tt arXiv:1212.5226}},
  \href{http://www.slac.stanford.edu/spires/find/hep/www?rawcmd=eprint+1212.5226}{[SPIRES]}.

\bibitem{Asaka:2005an}
T.~Asaka, S.~Blanchet, and M.~Shaposhnikov, {\it The {$\nu$MSM}, dark matter
  and neutrino masses},
  \href{http://dx.doi.org/10.1016/j.physletb.2005.09.070}{{\em Phys. Lett.}
  {\bf B631} (2005)  151--156},
  [\href{http://xxx.lanl.gov/abs/hep-ph/0503065}{{\tt hep-ph/0503065}}],
  \href{http://www.slac.stanford.edu/spires/find/hep/www?rawcmd=eprint+hep-ph\%2F0503065}{[SPRIES]}.

\bibitem{Asaka:2005pn}
T.~Asaka and M.~Shaposhnikov, {\it The {$\nu$MSM}, dark matter and baryon
  asymmetry of the universe},
  \href{http://dx.doi.org/10.1016/j.physletb.2005.06.020}{{\em Phys. Lett.}
  {\bf B620} (2005)  17--26},
  [\href{http://xxx.lanl.gov/abs/hep-ph/0505013}{{\tt hep-ph/0505013}}],
  \href{http://www.slac.stanford.edu/spires/find/hep/www?rawcmd=eprint+hep-ph\%2F0505013}{[SPRIES]}.

\bibitem{Kaiser:1994vs}
D.~I. Kaiser, {\it Primordial spectral indices from generalized einstein
  theories},  \href{http://dx.doi.org/10.1103/PhysRevD.52.4295}{{\em Phys.
  Rev.} {\bf D52} (1995)  4295--4306},
  [\href{http://xxx.lanl.gov/abs/astro-ph/9408044}{{\tt astro-ph/9408044}}],
  \href{http://www.slac.stanford.edu/spires/find/hep/www?rawcmd=eprint+astro-ph\%2F9408044}{[SPRIES]}.

\bibitem{Komatsu:1999mt}
E.~Komatsu and T.~Futamase, {\it {Complete constraints on a nonminimally
  coupled chaotic inflationary scenario from the cosmic microwave background}},
   \href{http://dx.doi.org/10.1103/PhysRevD.59.064029}{{\em Phys.Rev.} {\bf
  D59} (1999)  064029}, [\href{http://xxx.lanl.gov/abs/astro-ph/9901127}{{\tt
  astro-ph/9901127}}],
  \href{http://www.slac.stanford.edu/spires/find/hep/www?rawcmd=eprint+astro-ph\%2F9901127}{[SPRIES]}.

\bibitem{Liddle:1993fq}
A.~R. Liddle and D.~H. Lyth, {\it {The Cold dark matter density perturbation}},
   \href{http://dx.doi.org/10.1016/0370-1573(93)90114-S}{{\em Phys. Rept.} {\bf
  231} (1993)  1--105}, [\href{http://xxx.lanl.gov/abs/astro-ph/9303019}{{\tt
  astro-ph/9303019}}],
  \href{http://www.slac.stanford.edu/spires/find/hep/www?rawcmd=eprint+astro-ph\%2F9303019}{[SPRIES]}.

\bibitem{Gorbunov:2011zzc}
D.~S. Gorbunov and V.~A. Rubakov, {\em Introduction to the theory of the early
  universe: {C}osmological perturbations and inflationary theory}.
\newblock Hackensack, USA: World Scientific, 2011.

\bibitem{Linde:1983gd}
A.~D. Linde, {\it {Chaotic Inflation}},
  \href{http://dx.doi.org/10.1016/0370-2693(83)90837-7}{{\em Phys.Lett.} {\bf
  B129} (1983)  177--181},
  \href{http://www.slac.stanford.edu/spires/find/hep/www?rawcmd=j+PHLTA,B129,177}{[SPIRES]}.

\bibitem{Salopek:1992zg}
D.~Salopek, {\it {Consequences of the COBE satellite for the inflationary
  scenario}},  \href{http://dx.doi.org/10.1103/PhysRevLett.69.3602}{{\em
  Phys.Rev.Lett.} {\bf 69} (1992)  3602--3605},
  \href{http://www.slac.stanford.edu/spires/find/hep/www?rawcmd=j+PRLTA,69,3602}{[SPIRES]}.

\bibitem{Bezrukov:2007ep}
F.~L. Bezrukov and M.~Shaposhnikov, {\it The {S}tandard {M}odel {H}iggs boson
  as the inflaton},
  \href{http://dx.doi.org/10.1016/j.physletb.2007.11.072}{{\em Phys. Lett.}
  {\bf B659} (2008)  703--706}, [\href{http://xxx.lanl.gov/abs/0710.3755}{{\tt
  arXiv:0710.3755}}],
  \href{http://www.slac.stanford.edu/spires/find/hep/www?rawcmd=eprint+0710.3755}{[SPIRES]}.

\bibitem{Starobinsky:1983zz}
A.~Starobinsky, {\it {The Perturbation Spectrum Evolving from a Nonsingular
  Initially De-Sitter Cosmology and the Microwave Background Anisotropy}},
  {\em Sov.Astron.Lett.} {\bf 9} (1983)  302,
  \href{http://www.slac.stanford.edu/spires/find/hep/www?rawcmd=j+SALED,9,302}{[SPIRES]}.

\bibitem{Gorbunov:2010bn}
D.~Gorbunov and A.~Panin, {\it Scalaron the mighty: producing dark matter and
  baryon asymmetry at reheating},
  \href{http://dx.doi.org/10.1016/j.physletb.2011.04.067}{{\em Phys.Lett.} {\bf
  B700} (2011)  157--162}, [\href{http://xxx.lanl.gov/abs/1009.2448}{{\tt
  arXiv:1009.2448}}],
  \href{http://www.slac.stanford.edu/spires/find/hep/www?rawcmd=eprint+1009.2448}{[SPIRES]}.

\bibitem{Bezrukov:2008ut}
F.~Bezrukov, D.~Gorbunov, and M.~Shaposhnikov, {\it {On initial conditions for
  the Hot Big Bang}},
  \href{http://dx.doi.org/10.1088/1475-7516/2009/06/029}{{\em JCAP} {\bf 0906}
  (2009)  029}, [\href{http://xxx.lanl.gov/abs/0812.3622}{{\tt
  arXiv:0812.3622}}],
  \href{http://www.slac.stanford.edu/spires/find/hep/www?rawcmd=eprint+0812.3622}{[SPIRES]}.

\bibitem{Bezrukov:2011gp}
F.~Bezrukov and D.~Gorbunov, {\it {Distinguishing between $R^2$-inflation and
  Higgs-inflation}},
  \href{http://dx.doi.org/10.1016/j.physletb.2012.06.040}{{\em Phys.Lett.} {\bf
  B713} (2012)  365--368}, [\href{http://xxx.lanl.gov/abs/1111.4397}{{\tt
  arXiv:1111.4397}}],
  \href{http://www.slac.stanford.edu/spires/find/hep/www?rawcmd=eprint+1111.4397}{[SPIRES]}.

\bibitem{Gorbunov:2012ns}
D.~Gorbunov and A.~Tokareva, {\it {$R^2$-inflation with conformal SM Higgs
  field}},  \href{http://xxx.lanl.gov/abs/1212.4466}{{\tt arXiv:1212.4466}},
  \href{http://www.slac.stanford.edu/spires/find/hep/www?rawcmd=eprint+1212.4466}{[SPIRES]}.

\bibitem{Anisimov:2008qs}
A.~Anisimov, Y.~Bartocci, and F.~L. Bezrukov, {\it Inflaton mass in the
  {$\nu$MSM} inflation},
  \href{http://dx.doi.org/10.1016/j.physletb.2008.12.028}{{\em Phys. Lett.}
  {\bf B671} (2009)  211--215}, [\href{http://xxx.lanl.gov/abs/0809.1097}{{\tt
  arXiv:0809.1097}}],
  \href{http://www.slac.stanford.edu/spires/find/hep/www?rawcmd=eprint+0809.1097}{[SPIRES]}.

\bibitem{Beringer:1900zz}
{\bf Particle Data Group} Collaboration, J.~Beringer {\em et.~al.}, {\it
  {Review of Particle Physics (RPP)}},
  \href{http://dx.doi.org/10.1103/PhysRevD.86.010001}{{\em Phys.Rev.} {\bf D86}
  (2012)  010001},
  \href{http://www.slac.stanford.edu/spires/find/hep/www?rawcmd=j+PHRVA,D86,010001}{[SPIRES]}.

\bibitem{Bezrukov:2012sa}
F.~Bezrukov, M.~Y. Kalmykov, B.~A. Kniehl, and M.~Shaposhnikov, {\it Higgs
  boson mass and new physics},
  \href{http://dx.doi.org/10.1007/JHEP10(2012)140}{{\em JHEP} {\bf 1210} (2012)
   140}, [\href{http://xxx.lanl.gov/abs/1205.2893}{{\tt arXiv:1205.2893}}],
  \href{http://www.slac.stanford.edu/spires/find/hep/www?rawcmd=eprint+1205.2893}{[SPIRES]}.

\bibitem{Degrassi:2012ry}
G.~Degrassi, S.~Di~Vita, J.~Elias-Miro, J.~R. Espinosa, G.~F. Giudice,
  G.~Isidori, and A.~Strumia, {\it Higgs mass and vacuum stability in the
  {S}tandard {M}odel at {NNLO}},
  \href{http://dx.doi.org/10.1007/JHEP08(2012)098}{{\em JHEP} {\bf 1208} (2012)
   098}, [\href{http://xxx.lanl.gov/abs/1205.6497}{{\tt arXiv:1205.6497}}],
  \href{http://www.slac.stanford.edu/spires/find/hep/www?rawcmd=eprint+1205.6497}{[SPIRES]}.

\bibitem{Alekhin:2012py}
S.~Alekhin, A.~Djouadi, and S.~Moch, {\it {The top quark and Higgs boson masses
  and the stability of the electroweak vacuum}},
  \href{http://xxx.lanl.gov/abs/1207.0980}{{\tt arXiv:1207.0980}},
  \href{http://www.slac.stanford.edu/spires/find/hep/www?rawcmd=eprint+1207.0980}{[SPIRES]}.

\bibitem{Pilaftsis:1997jf}
A.~Pilaftsis, {\it {CP violation and baryogenesis due to heavy Majorana
  neutrinos}},  \href{http://dx.doi.org/10.1103/PhysRevD.56.5431}{{\em
  Phys.Rev.} {\bf D56} (1997)  5431--5451},
  [\href{http://xxx.lanl.gov/abs/hep-ph/9707235}{{\tt hep-ph/9707235}}],
  \href{http://www.slac.stanford.edu/spires/find/hep/www?rawcmd=eprint+hep-ph\%2F9707235}{[SPRIES]}.

\bibitem{Boyarsky:2009ix}
A.~Boyarsky, O.~Ruchayskiy, and M.~Shaposhnikov, {\it The role of sterile
  neutrinos in cosmology and astrophysics},
  \href{http://dx.doi.org/10.1146/annurev.nucl.010909.083654}{{\em
  Ann.Rev.Nucl.Part.Sci.} {\bf 59} (2009)  191--214},
  [\href{http://xxx.lanl.gov/abs/0901.0011}{{\tt arXiv:0901.0011}}],
  \href{http://www.slac.stanford.edu/spires/find/hep/www?rawcmd=eprint+0901.0011}{[SPIRES]}.

\bibitem{Canetti:2012kh}
L.~Canetti, M.~Drewes, T.~Frossard, and M.~Shaposhnikov, {\it {Dark Matter,
  Baryogenesis and Neutrino Oscillations from Right Handed Neutrinos}},
  \href{http://xxx.lanl.gov/abs/1208.4607}{{\tt arXiv:1208.4607}},
  \href{http://www.slac.stanford.edu/spires/find/hep/www?rawcmd=eprint+1208.4607}{[SPIRES]}.

\bibitem{Roy:2010xq}
A.~Roy and M.~Shaposhnikov, {\it Resonant production of the sterile neutrino
  dark matter and fine-tunings in the {$\nu$MSM}},
  \href{http://dx.doi.org/10.1103/PhysRevD.82.056014}{{\em Phys.Rev.} {\bf D82}
  (2010)  056014}, [\href{http://xxx.lanl.gov/abs/1006.4008}{{\tt
  arXiv:1006.4008}}],
  \href{http://www.slac.stanford.edu/spires/find/hep/www?rawcmd=eprint+1006.4008}{[SPIRES]}.

\end{thebibliography}\endgroup

\end{document}